\documentclass[twocolumn,english,prl,longbibliography]{revtex4-1}
\usepackage[T1]{fontenc}
\usepackage[latin9]{inputenc}
\usepackage{amsmath}
\usepackage{graphicx}
\usepackage{wasysym}

\makeatletter
\usepackage{hyperref}

\makeatother

\usepackage{babel}
\begin{document}
\title{Emergent Macroscopic Nonreciprocity from Identical Active Particles
via Spontaneous Symmetry Breaking}
\author{Wei-Chen Guo}
\affiliation{Institute for Theoretical Physics, School of Physics, South China
Normal University, Guangzhou 510006, China}
\affiliation{Key Laboratory of Atomic and Subatomic Structure and Quantum Control
(Ministry of Education), Guangdong Basic Research Center of Excellence
for Structure and Fundamental Interactions of Matter, School of Physics,
South China Normal University, Guangzhou 510006, China}
\affiliation{Guangdong Provincial Key Laboratory of Quantum Engineering and Quantum
Materials, Guangdong-Hong Kong Joint Laboratory of Quantum Matter,
South China Normal University, Guangzhou 510006, China}
\author{Zuo Wang}
\affiliation{Institute for Theoretical Physics, School of Physics, South China
Normal University, Guangzhou 510006, China}
\affiliation{Key Laboratory of Atomic and Subatomic Structure and Quantum Control
(Ministry of Education), Guangdong Basic Research Center of Excellence
for Structure and Fundamental Interactions of Matter, School of Physics,
South China Normal University, Guangzhou 510006, China}
\affiliation{Guangdong Provincial Key Laboratory of Quantum Engineering and Quantum
Materials, Guangdong-Hong Kong Joint Laboratory of Quantum Matter,
South China Normal University, Guangzhou 510006, China}
\author{Pei-Fang Wu}
\affiliation{Institute for Theoretical Physics, School of Physics, South China
Normal University, Guangzhou 510006, China}
\affiliation{Key Laboratory of Atomic and Subatomic Structure and Quantum Control
(Ministry of Education), Guangdong Basic Research Center of Excellence
for Structure and Fundamental Interactions of Matter, School of Physics,
South China Normal University, Guangzhou 510006, China}
\affiliation{Guangdong Provincial Key Laboratory of Quantum Engineering and Quantum
Materials, Guangdong-Hong Kong Joint Laboratory of Quantum Matter,
South China Normal University, Guangzhou 510006, China}
\author{Li-Jun Lang}
\email{ljlang@scnu.edu.cn}

\affiliation{Institute for Theoretical Physics, School of Physics, South China
Normal University, Guangzhou 510006, China}
\affiliation{Key Laboratory of Atomic and Subatomic Structure and Quantum Control
(Ministry of Education), Guangdong Basic Research Center of Excellence
for Structure and Fundamental Interactions of Matter, School of Physics,
South China Normal University, Guangzhou 510006, China}
\affiliation{Guangdong Provincial Key Laboratory of Quantum Engineering and Quantum
Materials, Guangdong-Hong Kong Joint Laboratory of Quantum Matter,
South China Normal University, Guangzhou 510006, China}
\author{Bao-Quan Ai}
\email{aibq@scnu.edu.cn}

\affiliation{Institute for Theoretical Physics, School of Physics, South China
Normal University, Guangzhou 510006, China}
\affiliation{Key Laboratory of Atomic and Subatomic Structure and Quantum Control
(Ministry of Education), Guangdong Basic Research Center of Excellence
for Structure and Fundamental Interactions of Matter, School of Physics,
South China Normal University, Guangzhou 510006, China}
\affiliation{Guangdong Provincial Key Laboratory of Quantum Engineering and Quantum
Materials, Guangdong-Hong Kong Joint Laboratory of Quantum Matter,
South China Normal University, Guangzhou 510006, China}
\author{Liang He}
\email{liang.he@scnu.edu.cn}

\affiliation{Institute for Theoretical Physics, School of Physics, South China
Normal University, Guangzhou 510006, China}
\affiliation{Key Laboratory of Atomic and Subatomic Structure and Quantum Control
(Ministry of Education), Guangdong Basic Research Center of Excellence
for Structure and Fundamental Interactions of Matter, School of Physics,
South China Normal University, Guangzhou 510006, China}
\affiliation{Guangdong Provincial Key Laboratory of Quantum Engineering and Quantum
Materials, Guangdong-Hong Kong Joint Laboratory of Quantum Matter,
South China Normal University, Guangzhou 510006, China}
\begin{abstract}
Nonreciprocity is known to generate a wide range of exotic phenomena
in multi-species many-body systems, where different species influence
one another through couplings that violate Newton\textquoteright s
third law. In contrast, in the absence of explicitly imposed macroscopic
nonreciprocal processes, single-species nonreciprocity---another
distinct form of nonreciprocity---typically plays only a limited
role in shaping macroscopic physics. Here, using a single-species
Vicsek model with a vision cone and extrinsic noise, we show that
spontaneous symmetry breaking (SSB) can dramatically enhance the macroscopic
consequences of microscopic single-species nonreciprocity. In the
ordered phase, this enhancement gives rise to an emergent macroscopic
nonreciprocity that induces the system of identical active particles
to admit an effective description with a \textquotedblleft two-species\textquotedblright{}
non-Hermitian structure. The resulting SSB-enhanced nonreciprocity
substantially promotes traveling-band formation and, more strikingly,
drives a novel real-space condensation of identical active particles,
characterized by a \textquotedblleft traveling line\textquotedblright{}
with vanishing longitudinal width. Our findings uncover a fundamental
mechanism by which microscopic single-species nonreciprocity can exert
strong macroscopic influences in complex systems.
\end{abstract}
\maketitle
\emph{Introduction.}---Nonreciprocity \citep{Fruchart_Nature_2021}
has attracted growing attention for its exceptional role in generating
exotic collective phenomena. In multi-species systems, nonreciprocal
interactions arise naturally when distinct particle types influence
one another in asymmetric ways---for example through chirality, quorum
sensing, or role differentiation such as leader--follower and predator--prey
dynamics \citep{Fruchart_Nature_2021,Klapp_Commun_Phys_2025,Saha_arXiv_2025,Tailleur_Nat_Commun_2023,Tailleur_JPhys_2024,DuanYu_PRL_2023,DuanYu_PRR_2025_QS,Vicsek_Nature_2010,Peruani_Nat_Phys_2022,Park_PRR_2020,ZhouTao_EPL_2015,Vicsek_NJP_2017,Lowen_EPL_2021,Lotka_JPhyschem_1910,Zhdankin_PRE_2010,Angelani_PRL_2012}.
These systems contain macroscopically distinct groups by construction,
nonreciprocal couplings are therefore built in at the collective level
and thus can directly shape macroscopic behavior. Such mechanisms
are widely studied in active and biological matter, particularly in
the multi-species Vicsek-type models with explicitly introduced leaders
or antagonistic agents \citep{ZhouTao_PRE_2009_Leader,Pearce_PRE_2016,Holubec_PRE_2022,Chatterjee_PRE_2023,Chatterjee_PRE_2025,Ihle_PRE_2025}.

Nonreciprocity can also arise in single-species systems. In models
with limited vision, such as vision-cone variants of the Ising, XY
and Vicsek models \citep{Durve_EurPhysJE_2018,Mishra_PRE_2018,Dadhichi_PRE_2020,ZhouTao_PRE_2009_Vision,Levis_JStat_2025,Loos_PRL_2023,Loos_JStatMech_2025},
interactions between identical particles may become transiently asymmetric.
This type of nonreciprocity is also ubiquitous in active and intelligent
matter, where interactions often depend on perception, communication,
or decision-making rules that need not satisfy Newton\textquoteright s
third law \citep{Lowen_arXiv_2025,Klapp_Nat_Commun_2025}. These systems
exhibit rich collective patterns---including nematic worms, milling
states, and other spatial structures \citep{Barberis_PRL_2016,Cheng_NJP_2016},
while the extent to which such macroscopic phenomena are directly
driven by the microscopic single-species nonreciprocity remains unclear
\citep{Vitelli_arXiv_2025}.

A key distinction between these two settings lies in the presence---or
absence---of stable macroscopic groups. In multi-species systems,
different particle populations exist a priori, allowing nonreciprocal
interactions to operate directly at the collective level \citep{Fruchart_Nature_2021,Klapp_Commun_Phys_2025,Saha_arXiv_2025,Tailleur_Nat_Commun_2023,Tailleur_JPhys_2024,DuanYu_PRL_2023,DuanYu_PRR_2025_QS,Vicsek_Nature_2010,Peruani_Nat_Phys_2022,Park_PRR_2020,ZhouTao_EPL_2015,Vicsek_NJP_2017,Lowen_EPL_2021,Lotka_JPhyschem_1910,Zhdankin_PRE_2010,Angelani_PRL_2012}.
In contrast, single-species systems consist of identical particles,
so any effective grouping must emerge dynamically rather than being
imposed. As a result, macroscopic effects of microscopic single-species
nonreciprocity are typically transient and may average out at large
scales, making their macroscopic impact less obvious \citep{Levis_JStat_2025,Loos_PRL_2023,Loos_JStatMech_2025,Lowen_arXiv_2025}.
This raises a fundamental question: under what circumstances can the
microscopic single-species nonreciprocity become macroscopically operative
and qualitatively reshape collective behaviors?

\begin{figure}
\begin{centering}
\includegraphics[width=3.3in]{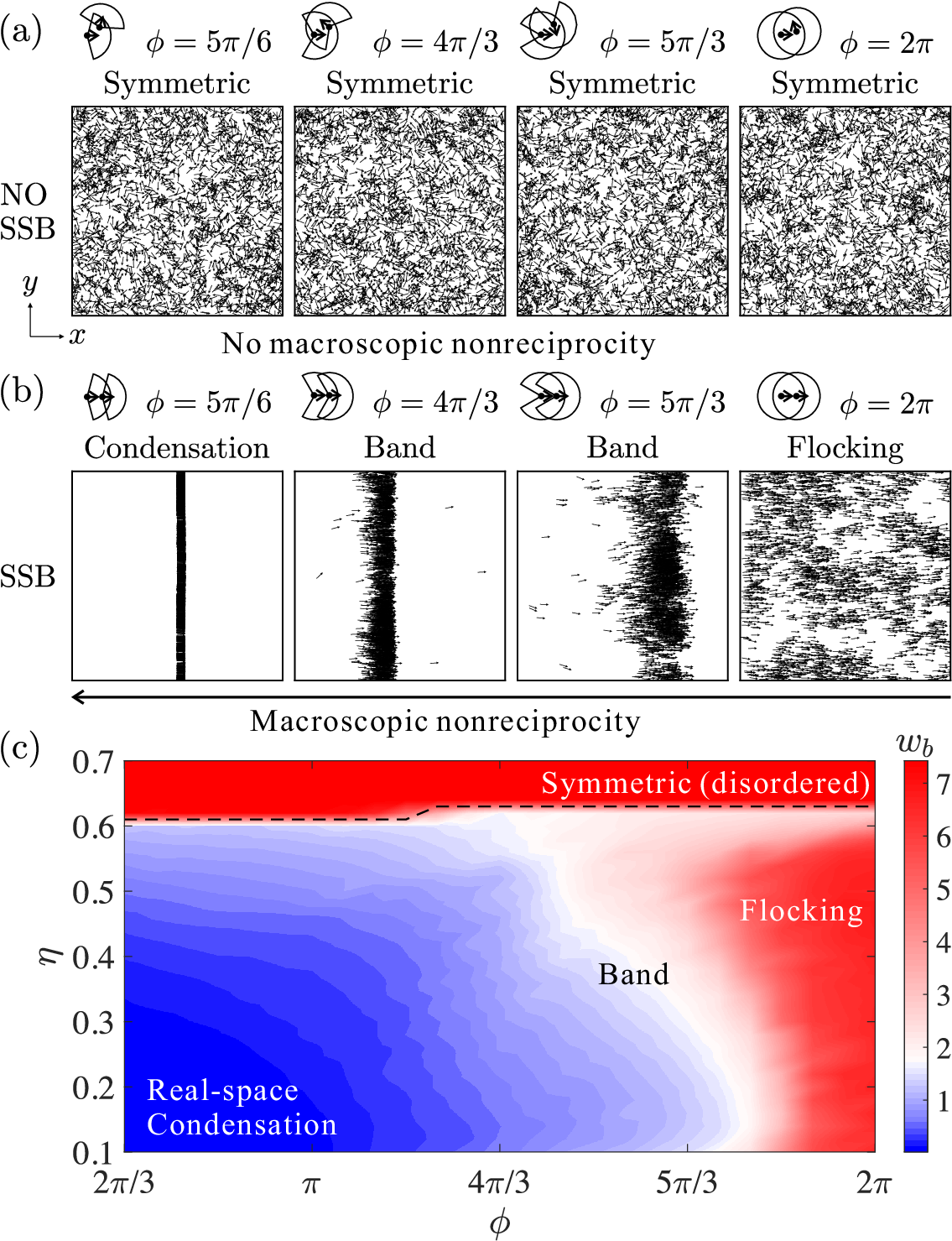}
\par\end{centering}
\caption{\label{fig:Illustration}(a) Typical steady-state configurations at
high noise, $\eta=0.75>\eta_{c}$, where the system is disordered.
In this regime, microscopic front--back nonreciprocity does not accumulate
macroscopically, and no qualitative differences are observed for different
view angles. (b) Typical steady-state configurations at low noise,
$\eta=0.1<\eta_{c}$, where SSB induces collective motion. In this
ordered phase, nonreciprocity is enhanced and produces pronounced
macroscopic effects. For sufficiently small view angle (leftmost panel),
the system exhibits real-space condensation, in which all particles
collapse to the same longitudinal position along the collective velocity.
(c) Distribution of the band width $w_{b}$ in the $(\eta,\phi)$
parameter space. The lower-left blue region corresponds to real-space
condensation with vanishing $w_{b}$. For $\eta<\eta_{c}$, decreasing
$\phi$ stabilizes traveling bands and eventually drives condensation.
In contrast, for $\eta>\eta_{c}$, varying $\phi$ produces no observable
effect (upper red region), demonstrating that macroscopic nonreciprocity
emerges only in the SSB phase. See text for details.}
\end{figure}

\begin{figure}
\begin{centering}
\includegraphics[width=3.3in]{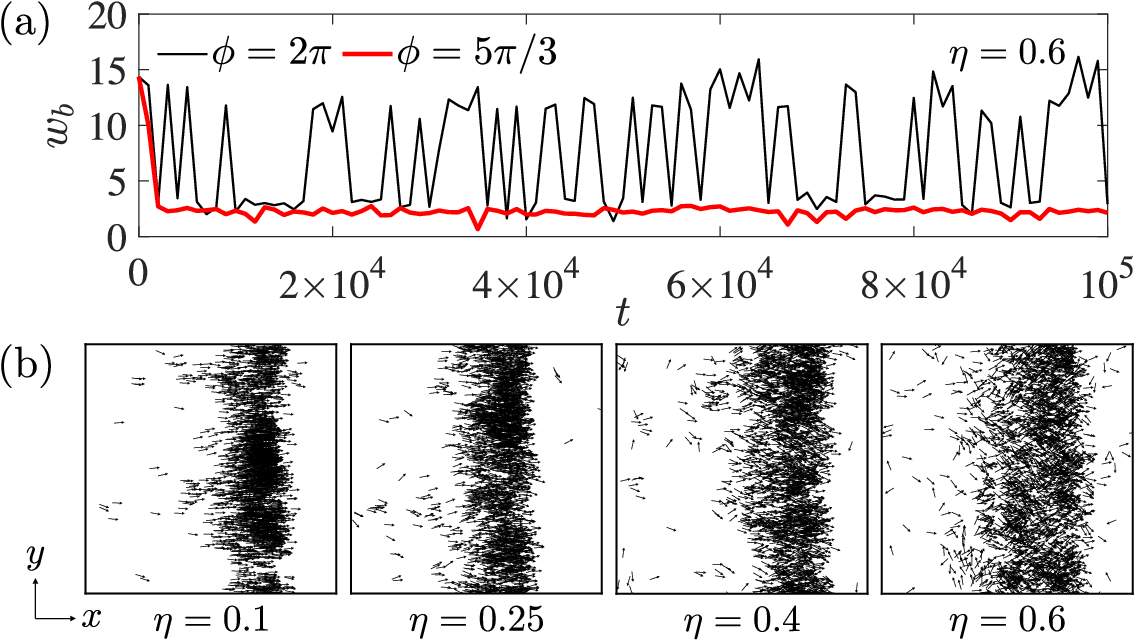}
\par\end{centering}
\caption{\label{fig:Trajectory}(a) Representative single-trajectory time evolution
of the band width $w_{b}$ at noise level $\eta=0.6$, slightly below
the flocking transition $\eta_{c}$ ($N=10^{4}$, $L=50$). For $\phi=5\pi/3$,
temporal fluctuations of $w_{b}$ are strongly suppressed, demonstrating
the stabilization of the traveling band by SSB-enhanced nonreciprocity.
(b) Typical steady-state configurations for $\phi=5\pi/3$, showing
robust traveling bands over a broad range of noise strengths ($N=2500$,
$L=25$). See text for details.}
\end{figure}

\begin{figure}
\begin{centering}
\includegraphics[width=3.3in]{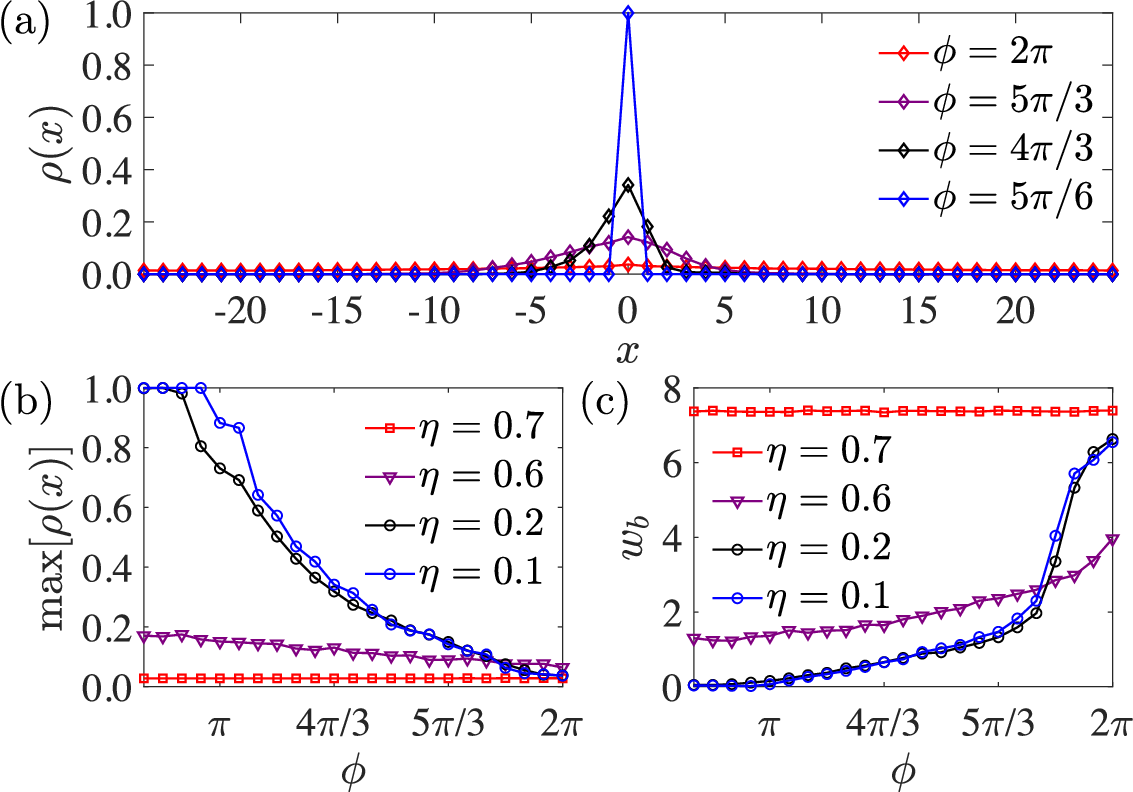}
\par\end{centering}
\caption{\label{fig:Main}(a) Normalized density profile $\rho(x)$ along the
collective velocity $\mathbf{v}_{c}$ for different view angles at
low noise $\eta=0.1$. For clarity, the peak of each distribution
is shifted to $x=0$. When $\phi$ becomes slightly smaller than $\pi$,
the profile collapses into a sharp peak, signaling real-space condensation.
(b) $\phi$-dependence of the peak value $\max[\rho(x)]$ at different
noise levels. For low noise ($\eta=0.1,\,0.2$), $\max[\rho(x)]$
reaches unity once $\phi\apprle\pi$, indicating complete longitudinal
condensation. (c) $\phi$-dependence of the band width $w_{b}$ at
different noise levels. For $\eta=0.1,\,0.2$, and $0.6<\eta_{c}$,
$w_{b}$ decreases as $\phi$ is reduced, reflecting enhanced longitudinal
compression. In contrast, for $\eta=0.7>\eta_{c}$, $w_{b}$ is essentially
independent of $\phi$, demonstrating that macroscopic nonreciprocity
is operative only in the SSB phase. See text for details.}
\end{figure}

Here, we address this question using a single-species Vicsek model
with a vision cone and extrinsic noise {[}see Eq.~(\ref{eq:Vicsek-Dynamics})
and the top of Fig.~\ref{fig:Illustration}(a){]}. A finite view
angle naturally generates front--back nonreciprocity: a particle
may lie within the visual field of another without the interaction
being mutual. This nonreciprocity is microscopic and transient, determined
by the instantaneous particle configuration, and thus fundamentally
distinct from permanent multi-species nonreciprocity. Although no
explicit macroscopic nonreciprocity is imposed a priori, we show that
its interplay with spontaneous symmetry breaking (SSB) at low noise
levels qualitatively reshapes the collective dynamics. Once global
alignment emerges, the microscopic front--back asymmetry becomes
coherently organized along the direction of the collective velocity,
effectively inducing nonreciprocal interactions between macroscopic
populations of front and back particles {[}see the top of Fig.~\ref{fig:Illustration}(b){]}.
Consequently, the system admits an effective description with an emergent
\textquotedblleft two-species\textquotedblright{} non-Hermitian structure,
despite being composed of identical particles {[}see Eq.~(\ref{eq:Fokker=002013Planck}){]}.
This SSB-enhanced macroscopic nonreciprocity substantially enlarges
the parameter regime in which traveling bands are stable {[}see Fig.~\ref{fig:Trajectory}
and Fig.~\ref{fig:Illustration}(c){]}, in sharp contrast to conventional
active matter systems, where the traveling band typically exists only
within a narrow noise interval slightly below the flocking transition
point $\eta_{c}$ \citep{Chate_PRE_2008,Chate_PRL_2004,Vicsek_PRL_1995,Toner_Ann_Phys_2005,Vicsek_Phys_Rep_2012,Volpe_RMP_2016,Chate_Annu_Rev_2020}.
More strikingly, we uncover a novel real-space condensation of identical
active particles, characterized by the formation of a \textquotedblleft traveling
line\textquotedblright{} with vanishing longitudinal width---an extreme
form of collective compression induced purely by SSB-enhanced nonreciprocity.
{[}see Fig.~\ref{fig:Main} and the leftmost panel of Fig.~\ref{fig:Illustration}(b){]}.

\emph{System and model with microscopic single-species nonreciprocity.}---We
consider a system of $N$ identical self-propelled particles moving
at a constant speed $v_{0}$ in a two-dimensional square domain of
size $L\times L$ with periodic boundary conditions. The particles
are subject to environmental fluctuations and interact via local alignment
within a fan-shaped neighborhood (including particle $i$ itself).
The collective dynamics is described by the following stochastic discrete-time
equation:

\begin{equation}
\boldsymbol{v}_{i}(t+\Delta t)=v_{0}\vartheta\left(\eta\mathcal{N}_{i}\boldsymbol{\xi}_{i}+\sum_{j\in U_{i}(r,\phi)}\boldsymbol{v}_{j}(t)\right).\label{eq:Vicsek-Dynamics}
\end{equation}
Here, $\Delta t$ is the time step and $\vartheta(\boldsymbol{w})=\boldsymbol{w}\slash\vert\boldsymbol{w}\vert$
is a normalization operator. $\mathcal{N}_{i}$ denotes the number
of neighbors of particle $i$. The interaction neighborhood $U_{i}(r,\phi)$
is a fan-shaped region centered at particle $i$, with interaction
radius $r$ and view angle $\phi$; its construction respects the
periodic boundary conditions. The random unit vector $\boldsymbol{\xi}_{i}$
represents vectorial extrinsic noise accounting for environmental
fluctuations, and $\eta\in[0,1]$ controls the noise strength. All
numerical results are obtained from direct simulations of Eq.~(\ref{eq:Vicsek-Dynamics}).
We set $r=1$ and $\Delta t=1$, so that length and time are measured
in units of $r$ and $\Delta t$, respectively. Unless otherwise stated,
we use $N=2500$, $L=25$, and $v_{0}=0.5$. Ensemble averages are
computed over $10^{2}$ independent stochastic trajectories.

To place our work in context, here we note that several variants of
the single-species Vicsek model with a vision cone have been explored
previously, often incorporating additional ingredients beyond the
intrinsic nonreciprocity induced by the vision cone. For example,
Ref.~\citep{Durve_EurPhysJE_2018} introduces communication time
delays; Ref.~\citep{Mishra_PRE_2018} considers random quenched rotators
as obstacles in the medium; and Ref.~\citep{Dadhichi_PRE_2020} includes
torques with rotational inertia. In contrast, the present model isolates
the minimal ingredient of single-species nonreciprocity, providing
a simple and generic framework.

\emph{SSB-enhanced macroscopic nonreciprocity and emergent ``two-species''
non-Hermitian structure.}---According to Eq.~(\ref{eq:Vicsek-Dynamics}),
a limited view angle $\phi<2\pi$ induces a front--back nonreciprocity
\citep{ShiXQ_PRE_2017} between two particles whenever one lies within
the field of view of the other but not vice versa {[}see the top of
Fig.~\ref{fig:Illustration}(a){]}. However, at high noise levels
$\eta>\eta_{c}$ (with $\eta_{c}\approx0.63$ marking the flocking
transition), the system remains disordered. Persistent random reorientations
prevent these microscopic nonreciprocal events from accumulating at
the macroscopic level. Thus, in the disordered phase, single-species
nonreciprocity produces no macroscopic effects. As shown in Fig.~\ref{fig:Illustration}(a),
steady-state configurations at high noise exhibit no qualitative differences
for different view angles.

The situation changes qualitatively in the presence of SSB. For $\eta<\eta_{c}$,
rotational symmetry is spontaneously broken and the system enters
the flocking phase, where a macroscopic fraction of particles moves
along the collective velocity $\mathbf{v}_{c}\equiv N^{-1}\sum_{i=1}^{N}\boldsymbol{v}_{i}$
\citep{Chate_PRE_2008,Chate_PRL_2004,Vicsek_PRL_1995}. In this ordered
phase, microscopic front--back nonreciprocity is no longer transient
but persists over long times, allowing its effects to accumulate macroscopically.
Along the direction of $\mathbf{v}_{c}$, particles that are more
frequently located at the front (\textquotedblleft front particles,\textquotedblright{}
$F$) exert a stronger net influence on those more frequently located
at the back (\textquotedblleft back particles,\textquotedblright{}
$B$) than vice versa {[}see the top of Fig.~\ref{fig:Illustration}(b){]}.
This dynamical imbalance effectively generates a nonreciprocal interaction
between two emergent \textquotedblleft species,\textquotedblright{}
$F$ and $B$.

To elucidate the consequences of this influence imbalance, we introduce
a minimal zero-dimensional effective model describing dynamical exchange
between $F$ and $B$. The model consists of two particle groups governed
by the transition processes $F\overset{\lambda_{FB}}{\rightarrow}B,\textrm{ and }F\overset{\lambda_{BF}}{\leftarrow}B,$
where $\lambda_{FB}$ ($\lambda_{BF}$) denotes the rate for a front
(back) particle to convert into a back (front) particle. The state
of the system at time $t$ is characterized by the probability distribution
$P(n_{B},n_{F};t)$, giving the probability of having $n_{F}$ front
particles and $n_{B}$ back particles. Using the standard procedure
\citep{Kamenev_Book_FTNEQS} (see Appendix A), we derive the corresponding
Fokker--Planck equation governing the evolution of $P(n_{B},n_{F};t)$:
\begin{align}
\partial_{t}P= & \left(e^{-\partial_{n_{B}}},e^{-\partial_{n_{F}}}\right)\Lambda\left(\begin{array}{c}
e^{\partial_{n_{B}}}n_{B}\\
e^{\partial_{n_{F}}}n_{F}
\end{array}\right)P,\label{eq:Fokker=002013Planck}
\end{align}
where $\Lambda\equiv\left(\begin{array}{cc}
-\lambda_{BF} & \lambda_{FB}\\
\lambda_{BF} & -\lambda_{FB}
\end{array}\right)$. The matrix $\Lambda$ is non-Hermitian whenever $\lambda_{BF}\neq\lambda_{FB}$.
In the flocking phase, such inequality naturally arises from the dynamical
influence-imbalance, suggesting that this non-Hermitian effective
description captures the essential macroscopic consequences of microscopic
single-species nonreciprocity in the presence of SSB.

The steady-state solution of Eq.~(\ref{eq:Fokker=002013Planck})
can be obtained analytically (see Appendix B). The average number
of front particles and its fluctuations are

\begin{equation}
\langle n_{F}\rangle=N\frac{\lambda_{BF}}{\lambda_{FB}+\lambda_{BF}},\,\Delta n_{F}=N\frac{\lambda_{BF}\lambda_{FB}}{(\lambda_{FB}+\lambda_{BF})^{2}},\label{eq:Steady-State-Solution}
\end{equation}
where $\Delta n_{F}\equiv\langle n_{F}^{2}\rangle-\langle n_{F}\rangle^{2}$.
A strong imbalance between $\lambda_{BF}$ and $\lambda_{FB}$ thus
drives particle accumulation at the front while suppressing number
fluctuations. This analysis predicts that SSB-enhanced single-species
nonreciprocity compresses the particle distribution along $\mathbf{v}_{c}$
and stabilizes it by reducing longitudinal number fluctuations.

\emph{Physical consequences of the SSB-enhanced nonreciprocity.}---The
above theoretical analysis points to potential macroscopic consequences
of SSB-enhanced nonreciprocity. A natural setting to explore such
effects is the traveling band, a paradigmatic collective structure
in Vicsek-type models. In reciprocal systems, the traveling band typically
exists only within a narrow noise interval slightly below the flocking
transition point $\eta_{c}$ \citep{Vicsek_PRL_1995,Chate_PRL_2004,Chate_PRE_2008,Toner_Ann_Phys_2005,Vicsek_Phys_Rep_2012,Volpe_RMP_2016,Chate_Annu_Rev_2020}.
Their dynamics involves recurrent disintegration and reformation \citep{Chate_PRE_2008},
accompanied by pronounced particle-number fluctuations inside the
band, which signal its intrinsic fragility. Since SSB-enhanced nonreciprocity
suppresses such fluctuations, we anticipate qualitative macroscopic
consequences---most notably, the stabilization of traveling bands
and an expansion of their parameter regime of existence.

Our numerical simulations support this expectation. For a given real-space
configuration, we define the band width as $w_{b}\equiv\sqrt{\langle x^{2}\rangle-\langle x\rangle^{2}}$,
where $\langle x^{2}\rangle=\int\rho(x)x^{2}dx$ and $\langle x\rangle=\int\rho(x)xdx$.
Here $\rho(x)=N^{-1}\int n(x,y)dy$ is the normalized density profile
along the collective direction $\mathbf{v}_{c}$; $x$ and $y$ denote
coordinates parallel and perpendicular to $\mathbf{v}_{c}$, respectively,
and $n(x,y)$ is the local particle density. Fig.~\ref{fig:Trajectory}(a)
shows the time evolution of $w_{b}$ along a single trajectory for
$\phi=2\pi$ and $5\pi/3$. Temporal fluctuations of $w_{b}$ are
clearly suppressed when $\phi=5\pi/3$, demonstrating that SSB-enhanced
nonreciprocity stabilizes the traveling band. Consistently, the band
persists over a broader parameter regime. As shown in Fig.~\ref{fig:Illustration}(b),
for $\phi=2\pi$ (reciprocal case) no traveling band is observed at
$\eta=0.1$. Upon decreasing $\phi$, the steady-state configurations
change qualitatively. The distribution of $w_{b}$ in the $(\eta,\phi)$
plane plane {[}see Fig.~\ref{fig:Illustration}(c){]} confirms that
SSB-enhanced nonreciprocity promotes band formation over a wide interval
of noise strength. For example, at $\phi=5\pi/3$, traveling bands
are observed for $0.1<\eta<0.62$ {[}see also Fig.~\ref{fig:Trajectory}(b){]}.

Equation (\ref{eq:Steady-State-Solution}) reveals that increasing
the imbalance between $\lambda_{BF}$ and $\lambda_{FB}$ drives progressive
particle accumulation in the front sector. In the extreme limit $\lambda_{FB}\ll\lambda_{BF}$,
the steady-state approaches $\langle n_{F}\rangle\rightarrow N$,
indicating that essentially all particles belong to the front group,
i.e., the effective non-Hermitian exchange dynamics collapses the
longitudinal particle distribution. Physically, this implies that
identical active particles self-organize to occupy the same longitudinal
position along the collective velocity $\mathbf{v}_{c}$. The density
profile along $\mathbf{v}_{c}$ therefore sharpens into a delta-like
peak, forming a narrow line perpendicular to $\mathbf{v}_{c}$. We
refer to this phenomenon as \emph{real-space condensation}, emphasizing
that it corresponds to a macroscopic occupation of the same longitudinal
position.

Fig.~\ref{fig:Main}(a) shows the normalized longitudinal density
profile $\rho(x)$ at fixed low noise $\eta=0.1$ for different view
angles. As $\phi$ decreases from $2\pi$, the distribution becomes
progressively narrower. When $\phi$ drops slightly below $\pi$,
the peak value $\max[\rho(x)]$ approaches unity, demonstrating that
all particles share the same longitudinal position {[}see also Fig.~\ref{fig:Illustration}(b){]}.
This observation directly confirms the mechanism revealed by the effective
non-Hermitian model: for $\phi<\pi$, front particles exert a negligible
alignment influence on back particles, leading to $\lambda_{FB}\ll\lambda_{BF}$
at low noise. The resulting strongly non-Hermitian exchange dynamics
funnels particles toward the front, producing complete longitudinal
collapse.

Interestingly, one notices that several collective patterns reported
in the literature may appear similar, including condensed drops \citep{Durve_EurPhysJE_2018},
traveling fibers \citep{Lowen_arXiv_2025}, nematic worms \citep{Barberis_PRL_2016},
spontaneous density segregation \citep{ShiXQ_PRL_2024}, or broad
traveling bands coexisting with a disordered background \citep{Lowen_arXiv_2025,Loos_PRL_2023,Loos_JStatMech_2025}.
However, we emphasize that the underlying mechanisms in those systems
are fundamentally different from the one considered here. These structures
typically emerge from additional ingredients, such as the combination
of a vision cone with time-delayed communication \citep{Durve_EurPhysJE_2018},
mechanical alignment torques \citep{Lowen_arXiv_2025}, position-based
attractive interactions \citep{Barberis_PRL_2016}, or lattice-coupled
bond dilution \citep{Loos_PRL_2023,Loos_JStatMech_2025}. In contrast,
the real-space condensation reported here arises purely from spontaneous-symmetry-breaking-enhanced
nonreciprocity within a single-species alignment interaction, without
invoking any additional mechanisms beyond the intrinsic nonreciprocity
induced by the vision cone.

To further reveal the influence of fluctuations on the real-space
condensation, we compute the $\phi$-dependence of $\max[\rho(x)]$
for different noise strengths. As shown in Fig.~\ref{fig:Main}(b,~c),
condensation occurs at low ($\eta=0.1$) and intermediate ($\eta=0.2$)
noise once $\phi<\pi$. By contrast, at higher noise ($\eta=0.6<\eta_{c}$),
no complete condensation emerges even for $\phi<\pi$. This can be
attribute to the fact that although SSB-enhanced nonreciprocity suppresses
$\lambda_{FB}$, environmental fluctuations generate a finite effective
back-to-front conversion rate, preventing complete collapse of the
longitudinal distribution. For $\eta>\eta_{c}$, varying $\phi$ produces
no noticeable effect, manifesting the crucial role played by SSB in
enhancing the macroscopic physical effects of the microscopic single-species
nonreciprocity. Finally, we emphasize that the above phenomena are
robust against variations in particle density, system size, and noise
type (see Appendix C), underscoring that they are generic and experimentally
accessible consequences of SSB-enhanced single-species nonreciprocity.

\emph{Conclusions.}---We have demonstrated that, in a single-species
Vicsek model with a vision cone and extrinsic noise, SSB can dramatically
amplify the macroscopic consequences of microscopic single-species
nonreciprocity. Although the underlying interactions involve identical
active particles, the SSB dynamically generates an effective \textquotedblleft two-species\textquotedblright{}
description governed by a non-Hermitian structure, thereby producing
emergent macroscopic nonreciprocity. We show that this SSB-enhanced
nonreciprocity stabilizes traveling bands and, more strikingly, drives
a novel real-space condensation of identical active particles. This
condensed state manifests as a \textquotedblleft traveling line\textquotedblright{}
with vanishing longitudinal width, representing an extreme form of
collective compression induced purely by dynamical nonreciprocity.

In contrast to multi-species systems---where nonreciprocal interactions
can directly shape macroscopic behavior---our results uncover a fundamental
mechanism by which comparatively weaker microscopic single-species
nonreciprocity becomes macroscopically operative through SSB. This
mechanism suggests a general route for amplifying hidden nonreciprocal
effects in nonequilibrium systems and may stimulate further exploration
in a broad range of active and collective matter settings, including
active spin flocks \citep{Chatterjee_PRE_2020,Caussin_PRL_2015,Solon_PRL_2013},
intelligent materials \citep{Kaspar_Nature_2021}, communicating active
particles \citep{Lowen_PNAS_2021}, and systems exhibiting herding
dynamics \citep{Boguna_PRE_2017}.
\begin{acknowledgments}
This work is supported by the NKRDPC (Grant No.~2022YFA1405304),
NSFC (Grants No.~11904109, No.~12075090, No.~12275089, and No.~12475036),
and the Guangdong Basic and Applied Basic Research Foundation (Grants
No.~2022A1515010449, No.~2023A1515012800, No.~2024A1515010188,
and No.~2024A1515012575), Guangdong Provincial Key Laboratory (Grant
No.~2020B1212060066).
\end{acknowledgments}

\appendix

\section*{End Matter}

\emph{Appendix A: Derivation of the Fokker--Planck equation of the
non-Hermitian effective description.}---In this appendix, for the
system in the presence of SSB, we present the derivation of the Fokker--Planck
equation of the non-Hermitian effective description. With the dynamical
rules $F\overset{\lambda_{FB}}{\rightarrow}B,\textrm{ and }F\overset{\lambda_{BF}}{\leftarrow}B$
in the effective model, the Fokker--Planck equation that determines
the time evolution of the probability distribution $P(n_{B},n_{F};t)$
can be obtained via a standard approach \citep{Kamenev_Book_FTNEQS}
by investigating the change of $P(n_{B},n_{F};t)$ with respect to
infinitesimal change of time. Concerning the probability of the configuration
$(n_{B},n_{F})$, its infinitesimal change $\delta P(n_{B},n_{F};t)$
with respect to the infinitesimal time change $\delta t$ consists
of two parts, namely an ``incoming-part'' 
\begin{align}
 & \delta t\lambda_{BF}(n_{B}+1)P(n_{B}+1,n_{F}-1;t)\nonumber \\
 & +\delta t\lambda_{FB}(n_{F}+1)P(n_{B}-1,n_{F}+1;t)
\end{align}
and an ``outgoing-part'' 
\begin{equation}
-\delta t\lambda_{BF}n_{B}P(n_{B},n_{F};t)-\delta t\lambda_{FB}n_{F}P(n_{B},n_{F};t).
\end{equation}
Collecting these two types of terms gives rise to 
\begin{align}
 & \delta P(n_{B},n_{F};t)\nonumber \\
= & -\delta t\lambda_{BF}n_{B}P(n_{B},n_{F};t)-\delta t\lambda_{FB}n_{F}P(n_{B},n_{F};t)\nonumber \\
 & +\delta t\lambda_{BF}(n_{B}+1)P(n_{B}+1,n_{F}-1;t)\nonumber \\
 & +\delta t\lambda_{FB}(n_{F}+1)P(n_{B}-1,n_{F}+1;t),
\end{align}
whose $\delta t\rightarrow0$ limit directly leads to the dynamical
equation for $P(n_{B},n_{F};t)$, i.e., the Fokker--Planck equation:
\begin{align}
 & \partial_{t}P(n_{B},n_{F};t)\nonumber \\
= & -\lambda_{BF}n_{B}P(n_{B},n_{F};t)-\lambda_{FB}n_{F}P(n_{B},n_{F};t)\nonumber \\
 & +\lambda_{BF}(n_{B}+1)P(n_{B}+1,n_{F}-1;t)\nonumber \\
 & +\lambda_{FB}(n_{F}+1)P(n_{B}-1,n_{F}+1;t),
\end{align}
One can further use the shift operator $e^{a\partial_{n}}$ ($a$
is a real number) to recast each term on the right-hand side of the
above equation, for instance, $\lambda_{BF}(n_{B}+1)P(n_{B}+1,n_{F}-1;t)=\lambda_{BF}e^{-\partial_{n_{F}}}e^{\partial_{n_{B}}}n_{B}P(n_{B},n_{F};t)$,
and then find the Fokker--Planck equation (\ref{eq:Fokker=002013Planck})
as presented in the main text. Besides, by using a standard coarse-graining-type
approach (see, e.g., a similar derivation presented in Ref.~\citep{Fruchart_Nature_2021}),
one could also derive Eq.~(\ref{eq:Fokker=002013Planck}) starting
from the microscopic dynamical equation (\ref{eq:Vicsek-Dynamics}),
and Eq.~(\ref{eq:Fokker=002013Planck}) corresponds to the leading-order
terms that arise after coarse-graining Eq.~(\ref{eq:Vicsek-Dynamics}).

\emph{Appendix B: Steady-state solution of the Fokker--Planck equation.}---In
this appendix, we present the steady-state solution of this Fokker--Planck
equation (\ref{eq:Fokker=002013Planck}) of the non-Hermitian effective
description of the system in the presence of SSB. It can be obtained
analytically via an algebraic approach. From the Fokker--Planck equation
(\ref{eq:Fokker=002013Planck}), one can see that the dynamics of
the probability distribution $P(n_{B},n_{F};t)$ is completely determined
by the four linear operators:
\begin{align}
 & \hat{B}^{\dagger}\equiv e^{-\partial_{n_{B}}},\,\hat{B}\equiv e^{\partial_{n_{B}}}n_{B},\\
 & \hat{F}^{\dagger}\equiv e^{-\partial_{n_{F}}},\,\hat{F}\equiv e^{\partial_{n_{F}}}n_{F}.
\end{align}
Interestingly, they satisfy the canonical commutation relation between
the bosonic creation annihilation operators, i.e., $[\hat{A^{\prime}},\hat{A}^{\dagger}]=\delta_{A^{\prime}A}$
with $A$ and $A^{\prime}$ being $B$ or $F$. This thus indicates
that one could solve Eq.~(\ref{eq:Fokker=002013Planck}) analytically
via an algebraic approach. To realize it, let us first diagonalize
the quadrature with respect to the linear operators (we set $\lambda_{FB}=\gamma$,
$\lambda_{BF}=1$, which is equivalent to choosing the time unit as
$\lambda_{BF}^{-1}$)
\begin{equation}
(\hat{B}^{\dagger},\hat{F}^{\dagger})\left(\begin{array}{cc}
-1 & \gamma\\
1 & -\gamma
\end{array}\right)\left(\begin{array}{c}
\hat{B}\\
\hat{F}
\end{array}\right),
\end{equation}
 by introducing a new set of operators 
\begin{align}
 & (\hat{\alpha}^{\dagger},\hat{\beta}^{\dagger})=(\hat{B}^{\dagger},\hat{F}^{\dagger})\left(\begin{array}{cc}
\gamma & -1\\
1 & 1
\end{array}\right),\\
 & \left(\begin{array}{c}
\hat{\alpha}\\
\hat{\beta}
\end{array}\right)=\frac{1}{1+\gamma}\left(\begin{array}{cc}
1 & 1\\
-1 & \gamma
\end{array}\right)\left(\begin{array}{c}
\hat{B}\\
\hat{F}
\end{array}\right),
\end{align}
satisfying the same commutation relation $[\hat{\alpha},\hat{\beta}^{\dagger}]=\delta_{\alpha\beta}.$
Using this new set of operators, one can reformulate the quadrature
in the diagonal form, i.e., 
\begin{align}
 & (\hat{B}^{\dagger},\hat{F}^{\dagger})\left(\begin{array}{cc}
-1 & \gamma\\
1 & -\gamma
\end{array}\right)\left(\begin{array}{c}
\hat{B}\\
\hat{F}
\end{array}\right)\nonumber \\
 & =(\hat{\alpha}^{\dagger},\hat{\beta}^{\dagger})\left(\begin{array}{cc}
0 & 0\\
0 & -1-\gamma
\end{array}\right)\left(\begin{array}{c}
\hat{\alpha}\\
\hat{\beta}
\end{array}\right).
\end{align}
We notice that the eigenvalue corresponding to $\hat{\alpha}$ ($\hat{\alpha}^{\dagger}$)
is zero, which corresponds to the conservation of the total probability
of the system. And the mode corresponding to $\hat{\beta}$ ($\hat{\beta}^{\dagger}$)
is a mode that decays exponentially with respect to time. Then, let
us look for the vacuum state for the operator $\hat{\alpha}$ and
$\hat{\beta}$. Since $\hat{\alpha}=(\hat{B}+\hat{F})\slash(1+\gamma)$
and $\hat{\beta}=(-\hat{B}+\gamma\hat{F})\slash(1+\gamma),$ the vacuum
state for $\hat{B}$ and $\hat{F}$ is also the vacuum state for $\hat{\alpha}$
and $\hat{\beta}$, i.e., 
\begin{equation}
P_{0}(n_{B},n_{F})=\delta_{n_{B},0}\delta_{n_{F},0}.\label{eq:Vacumm-State}
\end{equation}
With the vacuum state for $\hat{\alpha}$ and $\hat{\beta}$, all
the eigenstates of the system can be generated by acting $\hat{\alpha}^{\dagger}$
and $\hat{\beta}^{\dagger}$ successively on the vacuum state, similar
to the algebraic solution of the quantum harmonic oscillator. Therefore,
we can now directly obtain the time evolution of the probability distribution
$P(n_{B},n_{F};t)$ of the system from any given initial condition. 

Now let us consider a simplest but still non-trivial case, i.e., $P(n_{B},n_{F};t=0)=c_{B}\delta_{n_{B},1}\delta_{n_{F},0}+c_{F}\delta_{n_{B},0}\delta_{n_{F},1}$
with $c_{B,F}>0$ and $c_{B}+c_{F}=1$. Here, $P(n_{B},n_{F};t=0)$
can be expressed as a linear combination of $P_{1_{\alpha}}\equiv\hat{\alpha}^{\dagger}P_{0}(n_{B},n_{F})$
and $P_{1_{\beta}}\equiv\hat{\beta}^{\dagger}P_{0}(n_{B},n_{F})$,
i.e., 
\begin{equation}
P(n_{B},n_{F};t=0)=\frac{c_{B}+c_{F}}{1+\gamma}P_{1_{\alpha}}+\frac{\gamma c_{F}-c_{B}}{1+\gamma}P_{1_{\beta}}.
\end{equation}
According to the eigenvalues of $\alpha$ and $\beta$ modes, $P_{1\alpha}$
will keep unchanged while $P_{1_{\beta}}$ will decay exponentially
with respect to time during the evolution. So in the long time limit,
the steady-state probability distribution of the system assumes the
following form: 
\begin{align}
P(n_{B},n_{F};t=+\infty) & =\frac{1}{1+\gamma}P_{1_{\alpha}}\\
 & =\frac{1}{1+\gamma}\left(\gamma\delta_{n_{B},1}\delta_{n_{F},0}+\delta_{n_{B},0}\delta_{n_{F},1}\right).\nonumber 
\end{align}
From this example one can see that the steady-state solution of the
probability distribution can be expressed in an expansion involving
only the $\alpha$ modes whose expansion efficient is determined by
the initial probability distribution. For a system with fixed number
of total particles $N$, it is straightforward to obtain that the
steady-state probability distribution $P_{\mathrm{SS}}(n_{B},n_{F})$
is just $P_{N_{\alpha}}$, i.e.,

\begin{align}
P_{\mathrm{SS}}(n_{B},n_{F}) & =\frac{\left(\hat{\alpha}^{\dagger}\right)^{N}P_{0}(n_{B},n_{F})}{\sum_{n_{B,}n_{F}}\left(\hat{\alpha}^{\dagger}\right)^{N}P_{0}(n_{B},n_{F})}\label{eq:Steady-State-Distribution}\\
 & =\frac{N!}{n_{B}!n_{F}!}\left(\frac{\gamma}{1+\gamma}\right)^{n_{B}}\left(\frac{1}{1+\gamma}\right)^{n_{F}},\nonumber 
\end{align}
which is a binomial distribution. Utilizing this steady-state probability
distribution, one can directly calculate the key properties associated
with the number of front-particles or back-particles, and find the
results as presented in Eq.~(\ref{eq:Steady-State-Solution}) in
the main text.

\begin{figure}
\begin{centering}
\includegraphics[width=3.3in]{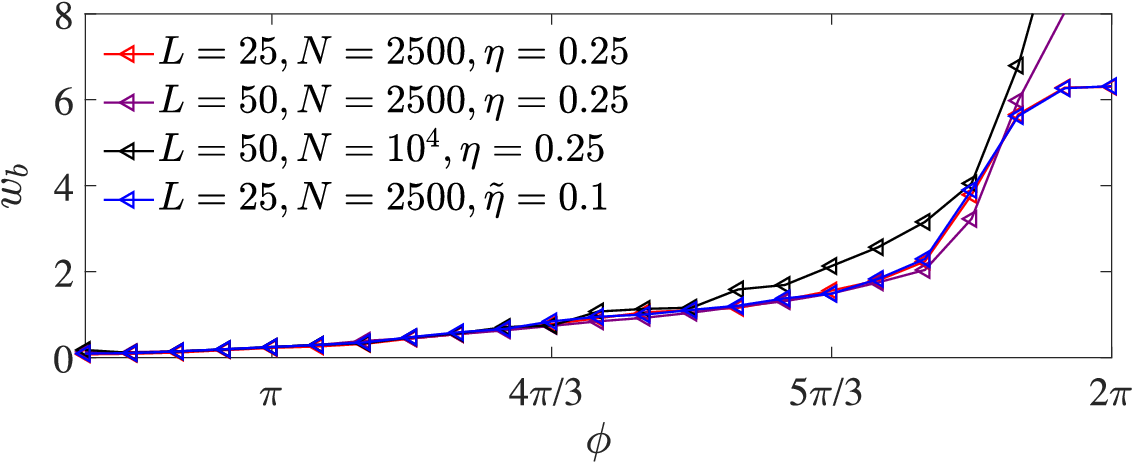}
\par\end{centering}
\caption{\label{fig:Other-properties}The view angle $\phi$ dependence of
the traveling band width $w_{b}$ in different cases. These results
show that the particle density and the system size do not assume strong
influences on the physical effects of the SSB-enhanced single-species
nonreciprocity. See text for more details.}
\end{figure}

\emph{Appendix C: Influences of particle density, system size, and
the type of noise.}---According to the dynamical equation (\ref{eq:Vicsek-Dynamics}),
the lower particle density $\bar{n}=N\slash L^{2}$ is expected to
make the system relatively more sensitive to influences from environmental
fluctuations. However, as one can see from the red and black curves
in Fig.~\ref{fig:Other-properties}, reducing the density $\bar{n}$
within a certain range do not prevent the real-space condensation
from existing. The purple curve in Fig.~\ref{fig:Other-properties}
indicates that finite-size effects do not impose strong influences
on the physical effects of the SSB-enhanced single-species nonreciprocity,
neither. Besides, when the SSB-enhanced nonreciprocity is strong enough
to driven the real-space condensation, we find that larger systems
(e.g., $N=10^{4}$) can accommodate several traveling ``lines''
at the same time without forming regular wave trains, which is similar
to the cases associated with the well-established traveling band \citep{Chate_PRE_2008}.
Furthermore, concerning the fluctuations, the extrinsic noise characterizes
influences from environmental fluctuations and assumes that active
particles are totally willing to be aligned with their neighbors.
There is another type of noise in the Vicsek model, namely, the intrinsic
noise (also known as ``angular'' noise) with $\tilde{\eta}\in[0,1]$
being the intrinsic noise level characterizing influences from the
``free will'' of the active particles \citep{Vicsek_PRL_1995}.
As one can see from the blue line in Fig.~\ref{fig:Other-properties},
the traveling band formation and real-space condensation can also
be found in this case. These results show that the particle density,
system size, and the type of noise do not assume strong influences
on the physical effects of the SSB-enhanced nonreciprocity, naturally
facilitating further direct experimental observations of such macroscopic
phenomena driven by the microscopic single-species nonreciprocity.

\end{document}